\documentclass[oldversion]{aa}

\usepackage{natbib}
\bibpunct{(}{)}{;}{a}{}{,}
\usepackage{txfonts}
\usepackage{graphicx}
\begin{document}

\title{CaII Infrared triplet line models in Classical T Tauri stars}

\author{R. Azevedo\inst{1,2,3}
 \and N. Calvet\inst{2}
  \and L. Hartmann\inst{2} 
  \and D.Folha\inst{1,4} 
  \and F.Gameiro\inst{1,3} 
  \and J. Muzerolle\inst{5}}

\institute{ Centro de Astrof\'isica da Universidade do Porto (CAUP), Rua das Estrelas, 4150-762 Porto, Portugal
\and Harvard-Smithsonian Center for Astrophysics, 60 Cambridge, MA 02138, USA 
\thanks {R. Azevedo as visiting student}
\and Departamento de Matem\'atica Aplicada da Faculdade de Ci\^encias da Universidade do Porto, Rua do Campo Alegre 687, 4169-007 Porto, Portugal
\and Instituto Superior de Ci\^{e}ncias da Sa\'{u}de - Norte, Rua Central da Gandra, 1317, 4585-116 GANDRA PRD, Portugal
\and Steward Observatory, University of Arizona, 933 North Cherry Avenue, Tucson, AZ 85721, USA} 

\date{Received / Accepted }

\abstract{We study the formation of the Calcium II infrared triplet lines 8498\AA, 8542\AA\ and 8662\AA, in the accreting magnetospheric flows of Classical T Tauri stars (CTTS), and present a grid of models for a large range of magnetospheric conditions. We apply our models to the interpretation of multi epoch observations of the CTTS DI Cep. We find that these lines form in the magnetospheric infall and that the variability of the CaII triplet lines in DI Cep can be explained in the context of changes in the mass accretion rate/temperature of the accretion column gas. }
\maketitle

\section{Introduction}

The concept of magnetically controlled accretion, introduced by \citet{Ghosh1977} for neutron stars, was applied to T Tauri stars by \citet{Uchida1985}, \citet{Camenzind1990}, \citet{Koenigl1991} and \citet{Shu1994}. From the interaction between the magnetic field lines and the disk, by transporting angular momentum outwards, it is possible to get the braking torque necessary to spin down T Tauri stars (TTS). Magnetocentrifugally driven winds can also remove angular momentum from the disk as it receives stellar angular momentum via the magnetospheric coupling \citep{Shu1994}. These theories can explain observations by \citet{Bouvier1993} and \citet{Edwards1993}, which show that Classical T Tauri Stars (CTTS), which accrete from disks, rotate slower than the non-accretors Weak T Tauri Stars (WTTS). However, \citet{Matt2004} argue that the disk braking torque may not be enough to spin down T Tauri stars at all.

In the context of CTTS \citet{Hartmann1994} and \citet{Muzerolle1998,Muzerolle2001} developed axisymmetric models of spectral lines formed in the magnetospheric accretion flow towards the star. They showed that the observed characteristics of the line profiles, such as (1) slightly blueshifted emission peaks, (2) blueward asymmetries, and (3) redshifted absorption components/inverse P Cygni profiles, could be well described by the models. The blue/UV excess emission coming from the high velocity material that shocks onto the star was also successfully modeled by \citet{Calvet1998} and \citet{Gullbring2000}. The magnetic field line intensities necessary to disrupt the material from the disk are consistent with measurements from \citet{Krull1999} and \citet{Krull2000}. 

Most modeling of lines supposedly formed in the magnetospheric accretion flow has been focused on hydrogen and sodium lines. But other tracers of the complex environment around TTS are also observed, such as the infrared triplet lines of CaII 8498\AA, 8542\AA, 8662\AA, or the HeI lines 10830\AA\ (actually a triplet) and 5876\AA, among others. We have started a modeling effort to simultaneously reproduce observations of these different lines, in order to restrict parameter space and to test the consistency of the model. These parameters include the mass accretion rate, the temperature of the gas and geometry. As a step in this direction we develop a study of the CaII infrared triplet.

The CaII infrared triplet line profiles from CTTS can be classified into three distinct groups, according to their shapes: only narrow component profiles (NC), only broad component profiles (BC), profiles with simultaneous NC and BC. When only the NC is present, the profiles are very similar to those of WTTS, which display enhanced narrow emission at the center of the photospheric absorption line \citep{Batalha1993,Batalha1996,Muzerolle1998}. BC emission lines have profiles extending to velocities in excess of 100 Km s$^{-1}$ \citep{Hamann1992,Batalha1996,Muzerolle1998}. The spectral samples presented in \citet{Hamann1992}; (sample 1) \citet{Batalha1996}; (sample 2) and \citet{Muzerolle1998}; (sample 3) give a full overview of these possibilities.

\section{Models}
\subsection{Line profile models}

We can obtain further insight into the infrared triplet line characteristics by comparing them to the more commonly studied $H_\alpha$ line. Using, for example, sample 1 and sample 2 one can compare the CaII 8542\AA\ profile with that of $H_\alpha$ for about 43 CTTS. The main conclusion of this qualitative analysis is that only for very few cases $H_\alpha$ and the CaII line exhibit similarities both in shape and extension of the line wings. The CaII line is in general less featured. Using other lines of the infrared triplet would lead to the same conclusions, since there are no significant structural differences among them. Profiles with blueshifted emission, slight redshifted absorption, and superposed blueshifted absorption can appear in both $H_\alpha$ and the Ca II infrared triplet lines.

 The CaII lines show a peculiar pattern, found in many T Tauri and Herbig AeBe stars \citep{Hamann1990}. If the 8498\AA, 8542\AA, 8662\AA\ lines were optically thin, their strength would follow the 1:9:5 ratio according to their $(gf)$ values of $0.055,0.49,0.27$. The observed ratios are usually close (but not equal) to the optically thick limit \citep{Herbig1980,Hamann1989,Hamann1990} which corresponds to the ratio 1:1:1. The more extreme observed peak intensity pattern is I$^{p}_{8498 \AA} >$ I$^{p}_{8542\AA} >$ I$^{p}_{8662\AA}$. The peak pattern more often observed just follows the relation I$^{p}_{8498\AA} >$ I$^{p}_{8662\AA}$, while the other ratios are typical,i.e., I$^{p}_{8542\AA} \ga $ I$^{p}_{8498\AA}$ and I$^{p}_{8542\AA} \ga $ I$^{p}_{8662\AA}$ \citep{Hamann1990,Hamann1992}. There is a tendency of the 8498\AA\ line to be the narrower line, specially when it becomes stronger, while the width of 8662\AA\ line is similar to that of the 8542\AA\ line \citep{Hamann1990,Hamann1992}.

A chromospheric origin of the NC present in CaII line profiles of CTTS has been proposed by \citet{Hamann1992}, \citet{Batalha1993} and \citet{Batalha1996}, with the enhancement relative to WTTS being driven by disk accretion phenomena acting indirectly on the chromosphere \citep{Batalha1993}. \citet{Basri1990} argue that the BC could be explained by the sum of emission arising in turbulent regions of different velocities, which would also account for the narrower than gaussian central peaks seen in many CTTS. \citet{Hamann1992} also suggested that the BC CaII lines formed by turbulence in an extended envelope. \citet{Gullbring1996} argued that the BC could be formed in the high density post shock region of the accretion flow. According to \citet{Calvet1998}, the temperature range immediately after the shock is around $10^5K - 10^6 K$ which is probably too high to leave enough CaII available. The velocity is typically of the order of 100 Km s$^{-1}$ in this region which is less than the velocity extent of the lines commonly observed. Deeper in the postshock where the temperatures are lower, the velocity has decreased even more. On the other hand, \citet{Muzerolle1998} showed that models based on magnetospheric accretion infall could reproduce the observed profile of the BC observed in BP Tau, and were consistent with the profiles of other species. 

In this paper we explore CaII triplet line models further, by spanning a larger range of magnetospheric physical conditions. Although there is evidence for non-axisymmetric accretion as a way to explain the observed variability and the hot spots \citep{Bertout1988,Bouvier1993,Muzerolle2001,Romanova2004}, we use an axisymmetric model for simplicity. We compare the model results with observations of DI Cep collected by \citet[]{Gameiro2004}, hereafter GFP05. We discuss which values of the parameters yield better fits, trying to constrain them and comparing to those derived from other work. We will also attempt to explain the observed line variability of DI Cep by changing the crucial magnetospheric accretion parameters,i.e. the mass accretion rate and temperature of the gas. 

\begin{figure}
\centering
\includegraphics[width=8.5cm,height=7.5cm]{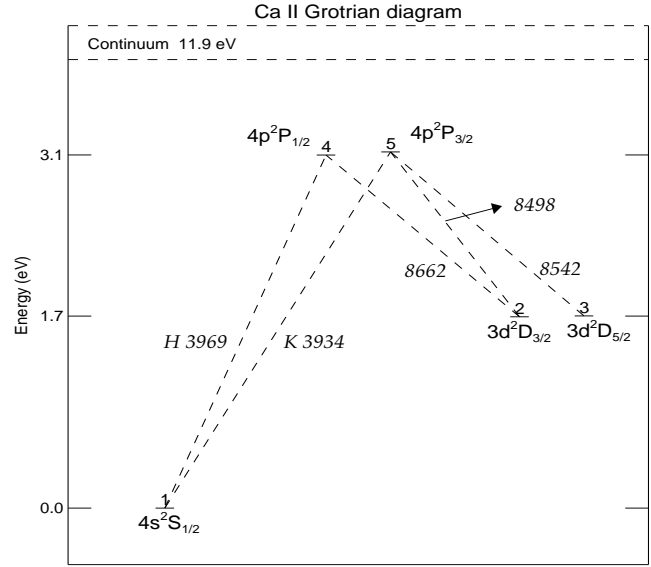}
\caption{\label{fig:grotrian} CaII Grotrian diagram showing the five energy levels plus continuum used in our models. The radiative transitions considered are represented by the dashed lines.}
\end{figure}

   We developed models of the CaII infrared triplet lines 8498\AA, 8542\AA\ and 8662\AA. The model atom consists of a five level ion plus continuum, with levels $4s~^2S_{1/2}$, $3d~^2D_{3/2,5/2}$ and $4p~^2P_{1/2,3/2}$ (see figure \ref{fig:grotrian}). The line profiles are computed based on the work of \citet{Hartmann1994}. We present in this section a brief review of the method.

   In the adopted model, the disk is truncated by the magnetic field, which has an assumed axisymmetric dipolar geometry and funnels the material towards the surface of the star where it arrives with near free fall velocity, merging onto the photosphere through a shock, simulated by a hot ring. The star is not rotating: as shown by \citet{Muzerolle2001} for the majority of the stars with slow rotational velocities, the effect is negligible in the final profiles. The numerical code CV, developed by \citet{Hartmann1994} to solve the radiative transfer in this kind of system, is used to calculate the source functions and radiation fields for the calcium lines. This code is based on the extended Sobolev approximation from \citet{Rybicki1978} which allows for a generic velocity flow geometry in relation to the standard Sobolev approximation. We then solve the statistical equilibrium equations to find the atomic level populations. The atomic data was collected from different sources: the collisional cross sections from PANDORA program; the $A_{ik}$,$f_{ik}$, energy level values were taken from the online Atomic Spectra Database V3.0 available at the US National Institute of Standards and Technology\footnote{http://physics.nist.gov/PhysRefData/ASD/index.html}, and the photoionization cross-sections from \citet{Shine1974}.
   
   Initially we set the input parameters of the model, i.e., the mass accretion rate ($\dot{M}_{acc}$), the geometry of the magnetosphere specified by the range of radii ($r_0-r_i$) which corresponds to the range where material is loaded from the disk into the magnetosphere, the temperature structure of the accretion flow characterized by its maximum value ($T_{gas}$) and the inclination ($i$). The effective temperature of the star ($T_{eff}$) has also to be specified as well as the mass ($M$) and radius ($R$). The temperature of the shock ($T_{r}$) can be derived from the previous parameters in light of the magnetospheric equations but we will assume it to be a 8000K blackbody, unless otherwise indicated.
   
   With these parameters set we use CV to generate the hydrogen and electron density structure across the magnetosphere. The temperature structure given the maximum value ($T_{gas}$) is calculated as described in \citet{Hartmann1994}.Considering the ionizing radiation field at a given point of the magnetosphere as the sum of a Planck contribution from the star and the ring multiplied by an adequate dilution factor,
\begin{equation}
\label{eq:Jtotal}
{J_{\nu}}=W_{star}B_{\nu\\}(T_{star})+W_{ring}B_{\nu}(T_{ring})
\end{equation}
with the $\overline{J}_{i}$ mean intensities of the 5 level system allowed transitions to be equal to the local Planck function, we can make an initial guess for the populations in each level, using the \cite{Mihalas} formalism, and a standard LU decomposition routine. We then use these populations to retrieve new $\overline{J}_{i}$ from CV, which we include in the statistical equilibrium equations, with the purpose of calculating new populations, repeating this procedure until the populations converge. Finally, the line profile is calculated summing all the emission from the resonant surfaces at a given frequency \citep[see][]{Hartmann1994}.

\begin{figure*}
\centering
\includegraphics[width=17cm]{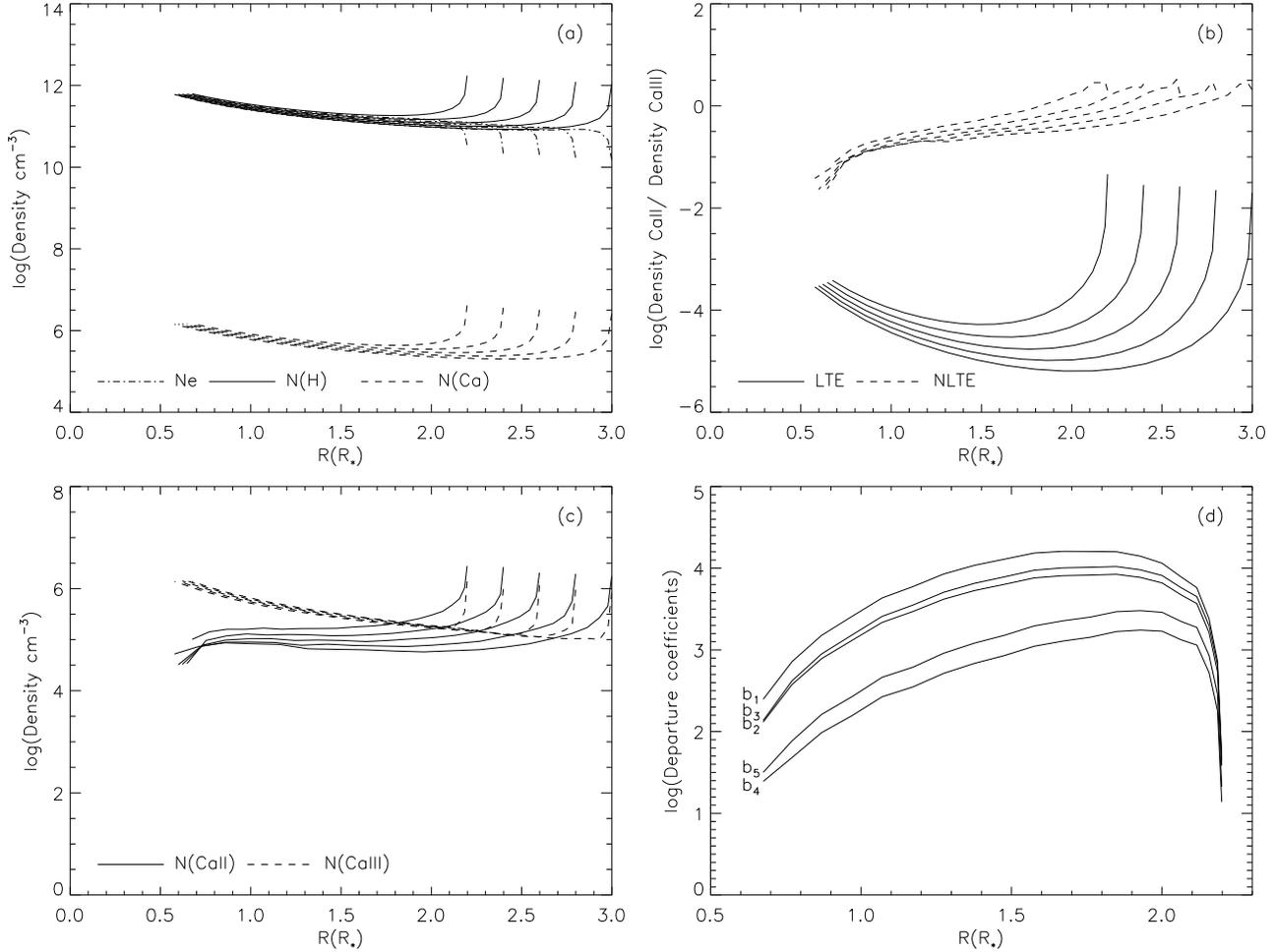}
\caption{\label{fig:compareabundances} Densities across the magnetosphere. (a) Total densities of H (solid line) and Ca (dashed line) across five streamlines, that intersect the disk at 2.2, 2.4, 2.6, 2.8 and $3R_\star$. $Ne$ is the electron density (dotted-dashed line). (b) Ratio between the density of CaII and CaIII in our NLTE models (dashed line) and in LTE (solid line). (c) Densities of CaII (solid line) and CaIII (dashed line). (d) Departure coefficients from LTE for the five levels in study. The departure coefficients are plotted for the streamline that intersects the disk at $2.2R_\star$. $R_\star=1.8R_{\odot},\; M_\star=1.42M_{\odot},\; T_\star=5400K,\; T_r=8000K,\; T_{gas}=12000K,\; log(\dot{M}_{acc})=-8.0,\; Geometry=2.2-3R_\star,\; i=60^o.$}
\end{figure*}

\begin{figure*}
\centering
\includegraphics[width=17cm]{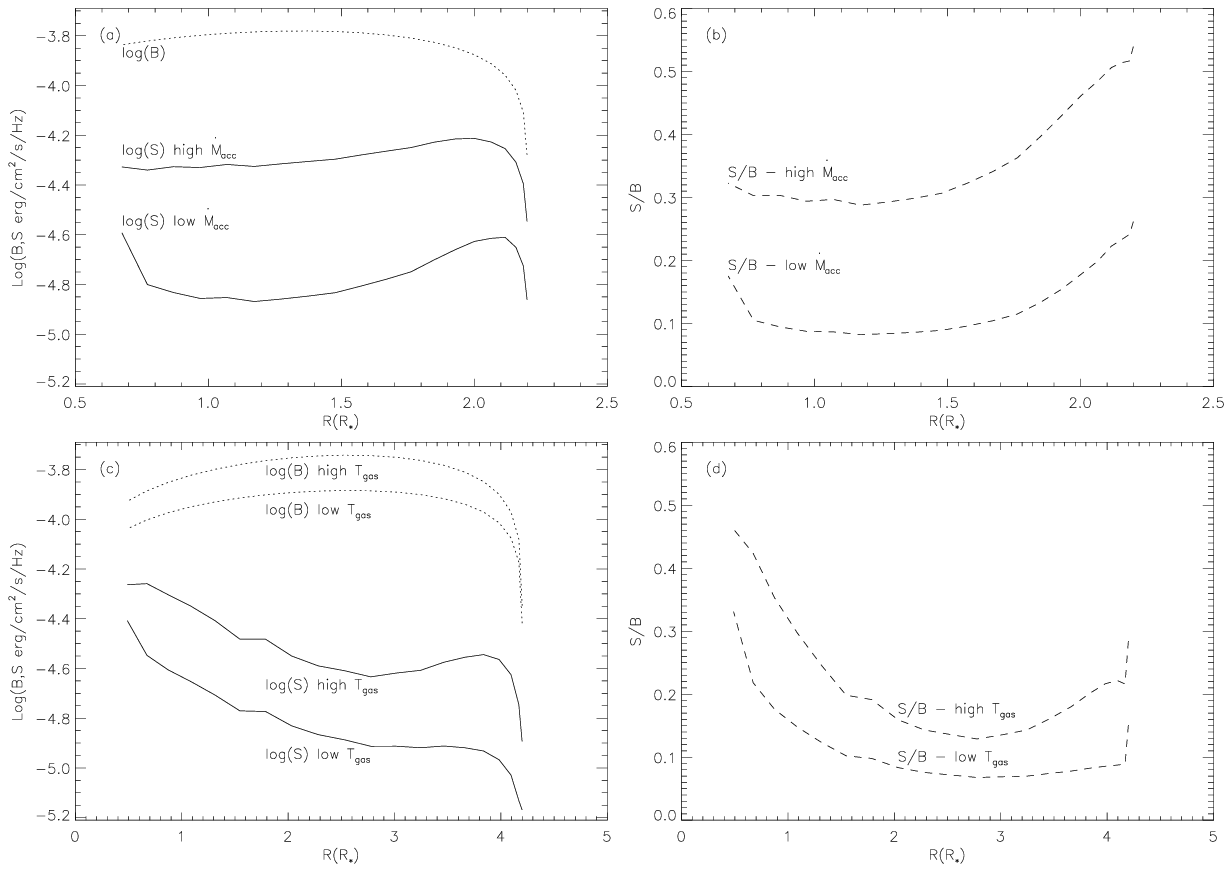}
\caption{\label{fig:sourceplanckit4by4} The Planck (B) and source (S) functions behavior for the 8662\AA\ line along a streamline, for different parameters. In (a) and (b) we use the geometry specified by the radii 2.2-3$R_\star$. B,S and S/B are plotted for the streamline that intersects the disk at 2.2$R_\star$. A high $\dot{M}_{acc}=2\times10^{-8}M_{\odot}yr^{-1}$ case and a low $\dot{M}_{acc}=1\times10^{-8}M_{\odot}yr^{-1}$ case are shown. In (c) and (d) we used the geometry specified by the radii 4.2-5$R_\star$. B,S and S/B are plotted for the streamline that intersects the disk at 4.2$R_\star$, in a high $T_{gas}=12000K$ and a lower $T_{gas}=10000K$ temperature regime. Other parameters as in Fig. \ref{fig:compareabundances}.}
\end{figure*}

In this section we explore the results from different input model parameters and initially describe some basic quantitative properties.

\subsubsection{Density}

\label{sec:Density}
In Fig. \ref{fig:compareabundances} we show the densities along the magnetosphere for a given set of parameters. The assumed abundance of Ca relative to H is 2.37$\times10^{-6}$ (similar to the solar system value in \citealp{Cox2000}). From setting the geometry and $\dot{M}_{acc}$ alone, we get in (a) a density behavior along the magnetosphere regulated by mass conservation. As the material is free falling, the increase inwards is due to the confinement of the material in a smaller area. As the velocity decreases rapidly near the disk and the area of the flow is more or less the same, the density rises. Comparing the density ratio of CaII/CaIII in LTE and in NLTE, shown in figure \ref{fig:compareabundances} (b), we can infer that the CaII density is much higher in the models than expected in LTE and the behavior is also different. The only continuum sources considered are the star and the accretion shock which have typical temperatures lower than those in the magnetosphere. Radiative recombination is therefore more important in the magnetosphere than photoionization in relation to a LTE case (defined as in equations 5-66 and 5-67 of \citealp{Mihalas}), so the CaII density increases by some orders of magnitude. Were the ratio between radiative recombination and photoionization constant throughout the magnetosphere and the variation of the ionization state with distance would be similar to LTE. The LTE behavior is just a consequence of the temperature profile (see Fig. \ref{fig:sourceplanckit4by4} Planck functions), and to a much less extent the electron density, so that is why CaIII is more important in the middle of the magnetosphere. In our NLTE case the ratio of CaII/CaIII is flatter than in LTE. This happens because the dilution factors are high and rise fast towards the star, and so photoionization rises fast too. Far from the star, about $\gtrsim 1.9R_\star$ for the streamline that intersects the disk at 2.2$R_\star$, the dilution factors decrease slowly and the gas temperature characteristic of radiative recombination decrease fast, so the rise in the ratio CaII/CaIII is not strong. Near the base of the magnetosphere, near the disk, the decrease of temperature is so fast that the ratio decreases, towards the LTE values. CaIII is the dominant state at $\lessapprox1.8R_\star$ and CaII the dominant state afterwards (panel c). However, the CaIII density rises near the disk, since photoionization and radiative recombination are more balanced again. Note that in a larger geometry than the one we show, a narrower flux tube near the star leads to higher total densities and electron densities. As a result, in panel (c) we would see the CaII density increase towards the star, following a behavior similar to that of H density in Fig. \ref{fig:compareabundances} (a), but more attenuated because of the photoionization rise. Panel (d) shows the departure coefficients from LTE for the five levels. The high values of the departure coefficients result from the increased CaII density in relation to LTE. The departure is higher for the lower levels and higher in the inner regions of the magnetosphere, which is evident in the continuum balance result shown in panel (b).

\subsubsection{Source Functions}
\label{sec:Sourcefunc}

Figure \ref{fig:sourceplanckit4by4} shows the behavior of the Planck and source function for the 8662\AA\ line along one streamline, for different parameters. In (a) we use a smaller geometry in two different mass accretion rate regimes and in (c) a larger geometry, where we also study the effects of different gas temperatures. The ratio of the source to the Planck functions are plotted in (b) and (d). In the low $\dot{M}_{acc}$ case in (a), we see a rise of the source function near the star. This happens because the collisional excitations from the metastable level $3d^2D_{3/2}$ to the level $4p^2P_{1/2}$ increase significantly due to the increase in the electron density. As we move further out, the source function decreases while the effect of the electron density in the collisional excitation is dominant. At a certain point the source function rises again because the CaII densities become rapidly higher and so the opacity of the H and 8662\AA\ lines increases. The escape probability decreases significantly and the upper level remains overpopulated in relation to the lower one. Near the end of the magnetosphere the electron density drops very fast (much lower temperatures), and so the strong decrease in the collisional excitation makes the source function decrease. The temperature drop effect is therefore not present in the ratio of the source function to the planck function plotted in panel (b).

In the high $\dot{M}_{acc}$ situation the increase in the CaII density makes escape in the line more difficult, and the increase in the electron density allows more collisional excitations. These two effects produce a source function that is more thermalized. Using panel (c), we can compare the high temperature case with the low mass accretion one of panel (a). Both cases only have different geometries. The rise of the source function towards the star is much more evident in the large geometry. This happens because the CaII density rises towards the star in the large geometry (see section \ref{sec:Density}). As a result we have more radiative trapping in relation to the smaller geometry case which makes the increased collisional excitation towards the star more evident. The source function for the lower gas temperature regime does not have a different behavior. As shown in panel (d), it is less thermalized than in the high temperature case, because of the lower electron densities, since hydrogen is less ionized.
 
 In Fig. \ref{fig:threesources} we plot the source functions for the five lines arising in the CaII system. We use the same parameters as in the low gas temperature case of Fig. \ref{fig:sourceplanckit4by4} (c) with the exception of a narrower geometry: 4.8-5$R_\star$. Note that the 8662\AA\ line is more thermalized than in Fig. \ref{fig:sourceplanckit4by4} (c), because narrower geometries imply higher densities. One can see that the source functions for the five lines arising in the CaII system are close to each other (valid to other sets of parameters). Near the star and close to the disk, the H and K source functions approach each other as well as the IR triplet lines. In other regions the source functions diverge, because the lines are less thermalized in the inner regions of the magnetosphere.  The 8498\AA\ source function stays close to the 8542\AA\ source function since the lines share a common upper level. We also note the relation $S_{42}/S_{41}\approx S_{52}/S_{51}$ \citep{Shine1974,Zirker1965}, which means that the changes in the H and K lines source functions affect similarly the 8662\AA\ and 8498\AA\ line source functions. 

\subsubsection{Line profiles}
\subsection{Exploration of model parameters}

\begin{figure}
	\centering
	\resizebox{\hsize}{!}{\includegraphics{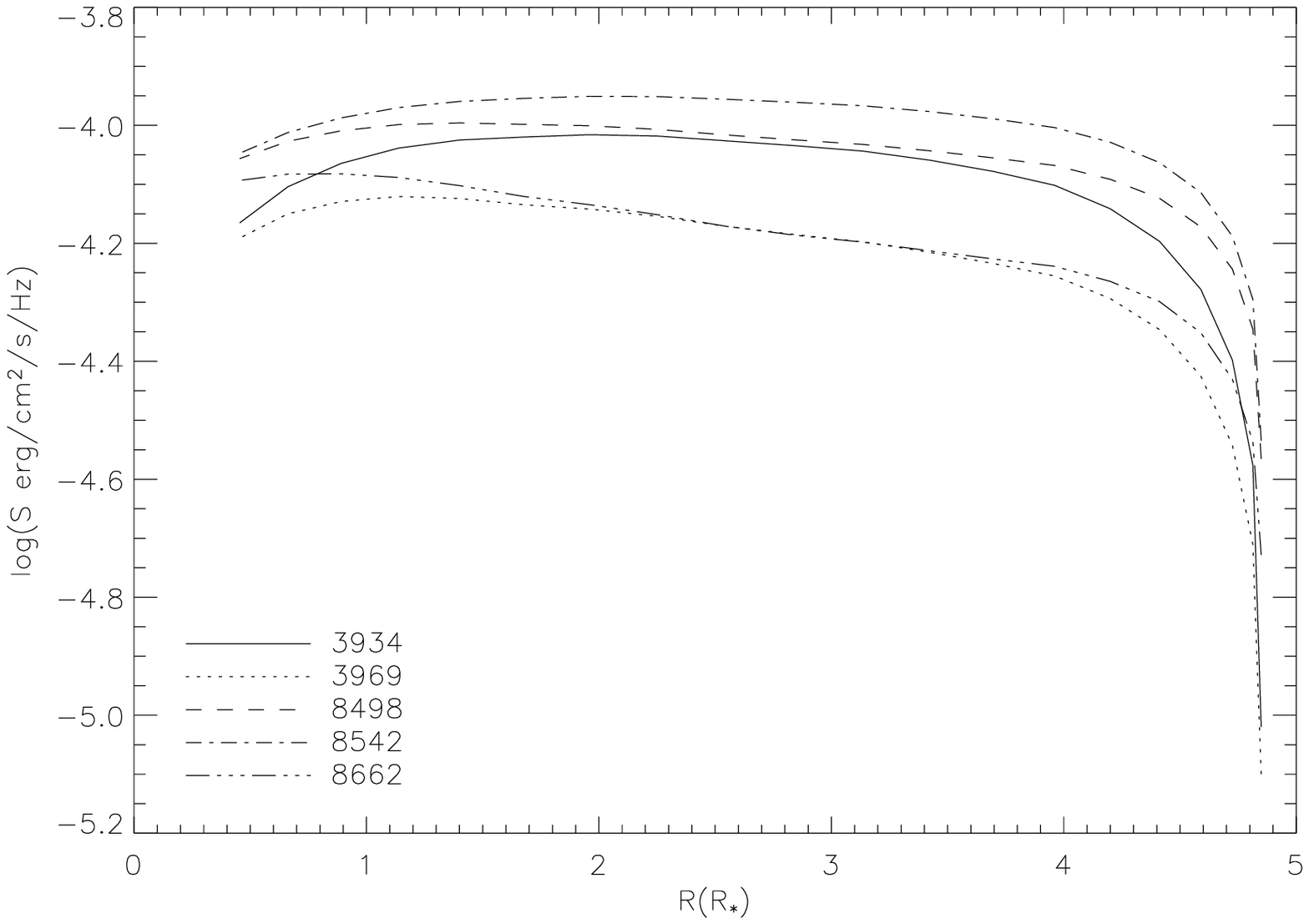}}
	\caption{\label{fig:threesources} Source functions for the five transitions considered in the CaII system, along the streamline that intersects the disk at 4.8$R_\star$. In the extremities of the magnetosphere the H and K line source functions approach each other as well as the infrared triplet source functions (see text). We use the same parameters as in the low gas temperature case of Fig. \ref{fig:sourceplanckit4by4} (c) with the exception of the Geometry=4.8-5$R_\star$. Note that a narrower geometry has higher densities, so the line source function 8662\AA\ is more thermalized than in Fig.\ref{fig:sourceplanckit4by4} (c).}
\end{figure}

\begin{figure*}
	\centering
	\includegraphics[width=12cm]{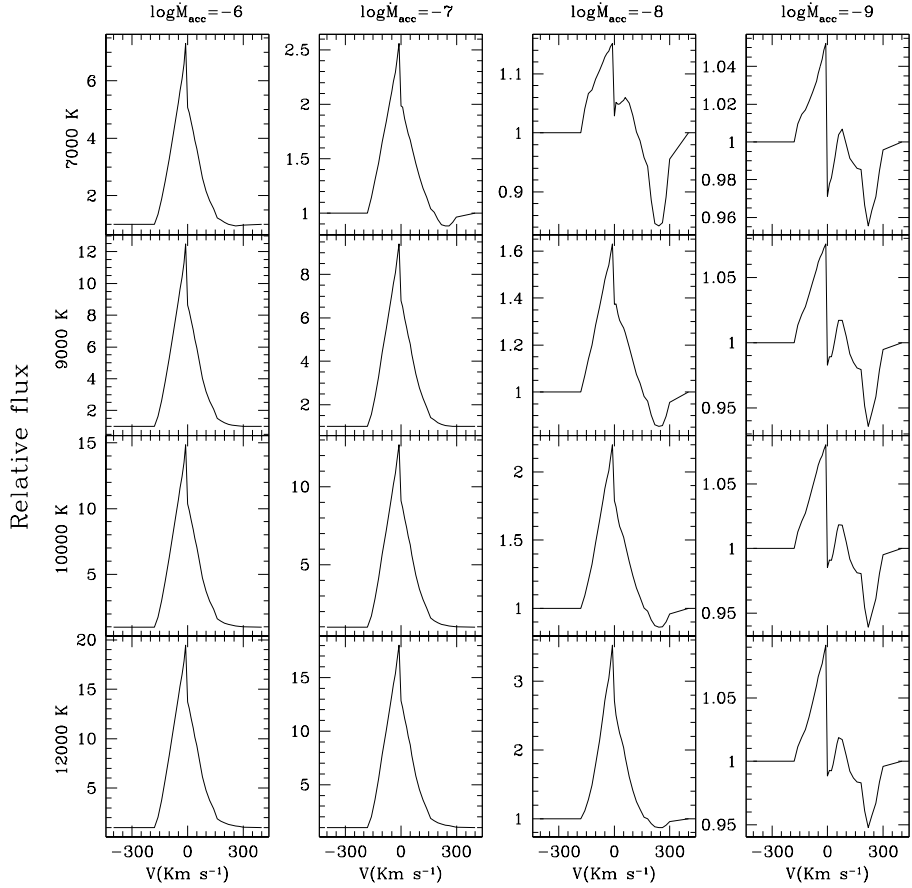}
	\caption[Normalized 8662\AA\ line profiles for different mass accretion rates and high temperatures of the gas.]{\label{fig:gridmasstemphightemps} Normalized 8662\AA\ line profiles for different mass accretion rates and high temperatures of the gas. Fixed parameters: $R_\star=1.9R_{\odot},\; M_\star=0.8M_{\odot},\; T_\star=4000K,\; T_r=8000K,\; Geometry=2.2-3R_\star,\; i=60^o.$}
\end{figure*}

In this section, we show line profiles for a range of mass accretion rates, temperatures and geometries. The behaviors we find are similar to those indicated in \citet{Hartmann1994,Muzerolle2001}, although we do not consider the effects resultant from including line broadening mechanisms and continuum opacity. We neglect radiative, Van der Waals and Stark broadening. Without a proper ray-by-ray treatment it is difficult make a simple argument. However, the broadening will be much less important than for example in $H_{\alpha}$ or $H_{\beta}$. For the exploratory nature of this paper therefore the assumption is adequate. As shown in Fig. \ref{fig:gridmasstemphightemps}, fixing the temperature, for decreasing mass accretion rates, we find less emission due to the decrease of densities. Strong redshifted absorption appears because of the growing importance of the background continuum relatively to the line source function \citep{Muzerolle2001}. When fixing the mass accretion rate and increasing the temperature the lines become more thermalized, resulting in higher line emission, and in more symmetric profiles. For the low mass accretion rates and low temperatures the model profiles do not have observational correspondence. For the highest mass accretion rates and temperatures, the line does not go into absorption as in \citet{Muzerolle2001} because of no continuum opacity treatment in our work.

\begin{figure*}
	\centering
	\includegraphics[width=12cm]{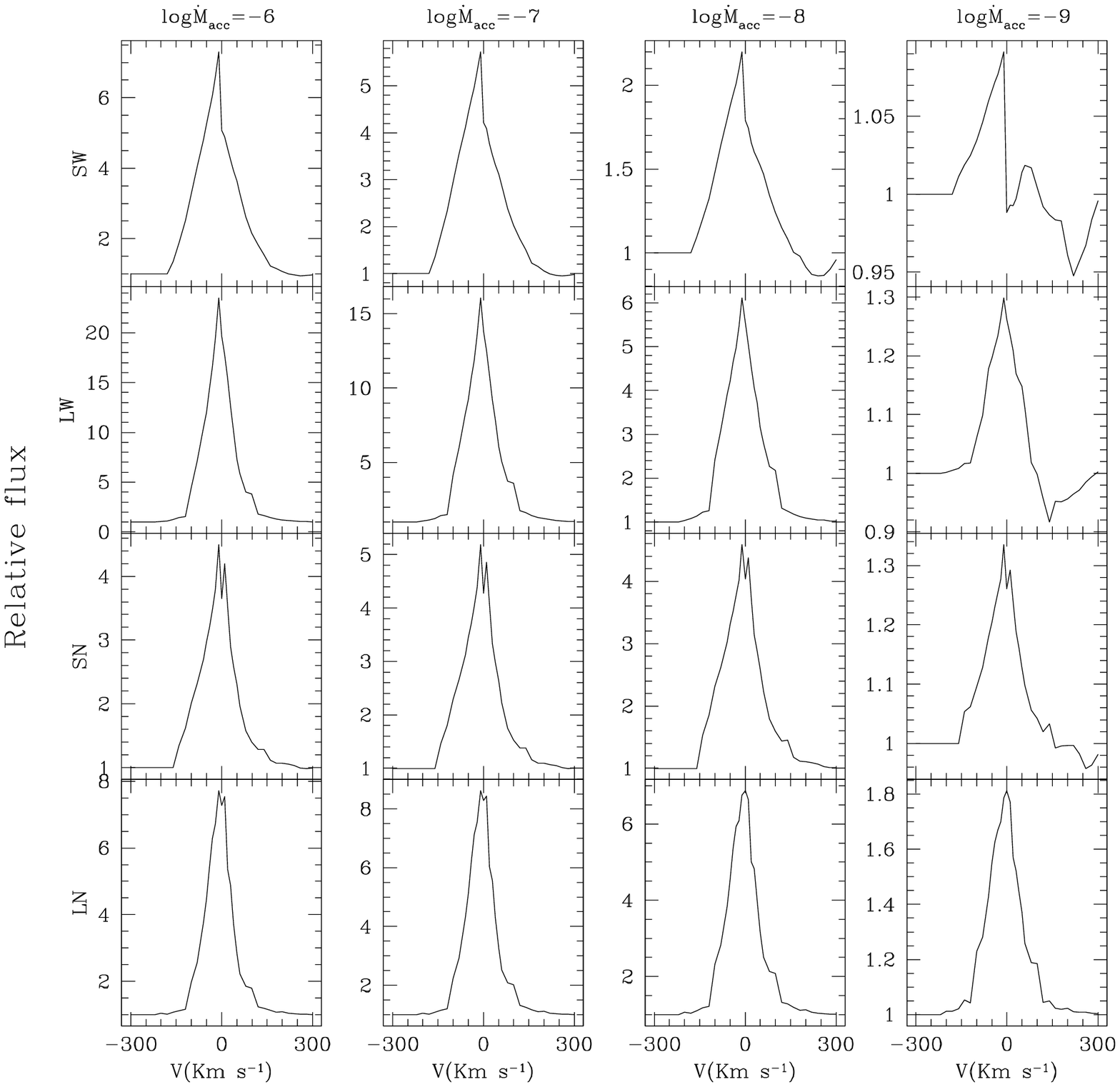}
	\caption[Normalized 8662\AA\ line profiles for different mass accretion rates and geometries of the gas.]{\label{fig:gridmasssizedicep} Normalized 8622\AA\ line profiles for different mass accretion rates and geometries of the gas. For better illustration, we used different $T_{gas}$. Log$\dot{M}_{acc}$=-6 ($T_{gas}$=7000K), Log$\dot{M}_{acc}$=-7 ($T_{gas}$=8000K), Log$\dot{M}_{acc}$=-8 ($T_{gas}$=10000K), Log$\dot{M}_{acc}$=-9 ($T_{gas}$=12000K). Fixed parameters: $R_\star=1.9R_{\odot},\; M_\star=0.8M_{\odot},\; T_\star=4000K,\; T_r=8000K,\; i=60^o$. }
\end{figure*}
\begin{figure*}
	\centering
	\includegraphics[width=12cm]{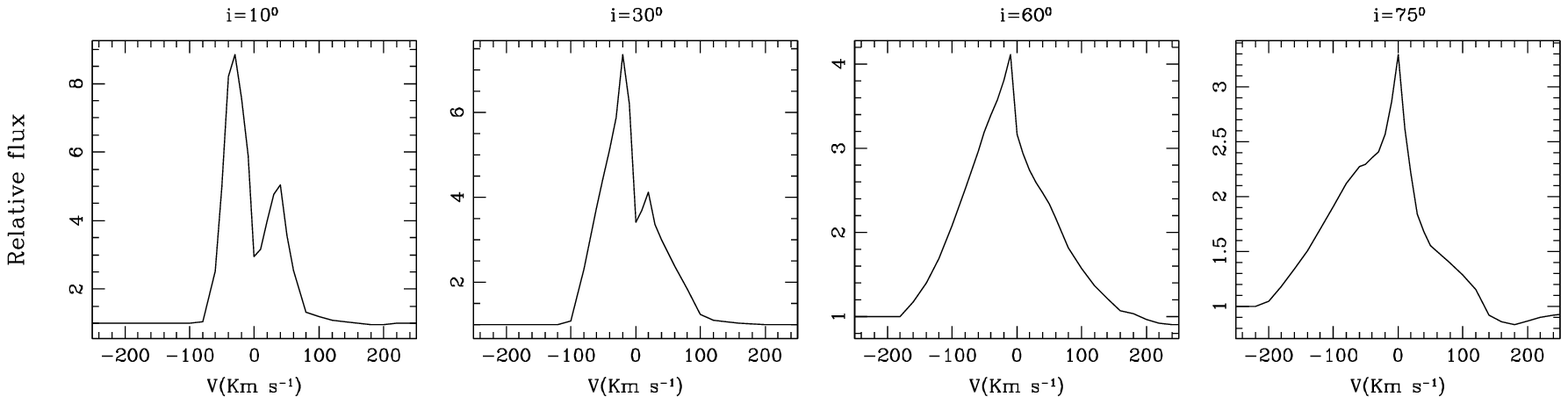}
	\caption[Normalized 8662\AA\ line profiles for different inclinations.]{\label{fig:gridinclitemp} Normalized 8662\AA\ line profiles for different inclinations. Fixed parameters: $T_{gas}=12000K,\; R_\star=1.9R_{\odot},\; M_\star=0.8M_{\odot},\; T_\star=4000K,\; T_r=7600K,\; log(\dot{M}_{acc})=-8,\; Geometry=2.2-3R_\star$}
\end{figure*}

In Fig. \ref{fig:gridmasssizedicep} we show profiles for various geometries and values of $\dot{M}_{acc}$. For better illustration, we use a high $T_{gas}$ for low $\dot{M}_{acc}$. We see more emission in the larger geometries than in the smaller ones, specially near the line center. The profiles are also narrower and more symmetric. In a large geometry, for the same inclination angle ($60^o$), the line of sight crosses much less higher velocity material resulting in narrower profiles. Large geometries have higher emitting volumes, and due to the more extended flow and less curvature, the line of sight for this inclination crosses more surfaces where the velocity projected in the observer's direction is nearly zero, resulting in increased emission, mainly near the line center. The increased symmetry results from less occultation effects from the star and the disk \citep{Muzerolle2001}. The difference in intensity when we compare wide and narrow geometries resides in the fact that in the latter case we have higher densities across each line of sight, since the accretion rate is the same and we confine the material in a smaller volume. In some of the profiles there seems to be a NC on top of a BC however they are wider than the typical NC widths found in CTTS of $\leq 50Km\;s^{-1}$ \citep{Batalha1996,Beristain2001}.

In Fig. \ref{fig:gridinclitemp} we explore the inclination dependence. For high inclinations the profile is broader since our line of sight crosses more high velocity material. For low inclinations, a deep absorption is seen near zero velocity that comes from the fact that our line of sight is parallel to the flow near the disk which is assumed to have a startup velocity of 10 km s$^{-1}$. The behavior is the same as in the hydrogen profiles modeled by \citet{Hartmann1994}, remarking that the broadening effects can smooth slightly the profile structure \citep{Muzerolle2001}. The apparent NC on top of a BC for $i=75^o$ comes from emission near to the disk. It has a base width of the order of the typical maximum of 50 $Km\;s^{-1}$ of NC in CTTS \citep{Beristain2001,Batalha1996}. However, the presence of an apparent narrow component on our model profiles is very dependent on the inclination. Such a geometrical effect does not seem to be observed. Additionaly, unlike the observed narrow components which show frequently unshifted or redshifted line centers \citep{Beristain2001,Batalha1996}, this feature has a slightly blueshifted center of emission.

The equivalent width ratios between the different CaII triplet model line profiles tend to be roughly around the optically thick regime, usually with the 8542\AA\ line slightly enhanced relative to its neighbours.

It should be kept in mind that the Sobolev approximation close to zero velocity does not hold, which is the main reason behind the sharp peaks in the line center. In future work, we will carry out a more adequate proper escape probability treatment. Broadening mechanisms and the effects of rotation will also be explored.

\section{Model application to DI Cep observations}
\begin{figure*}
\centering
\includegraphics[width=17cm]{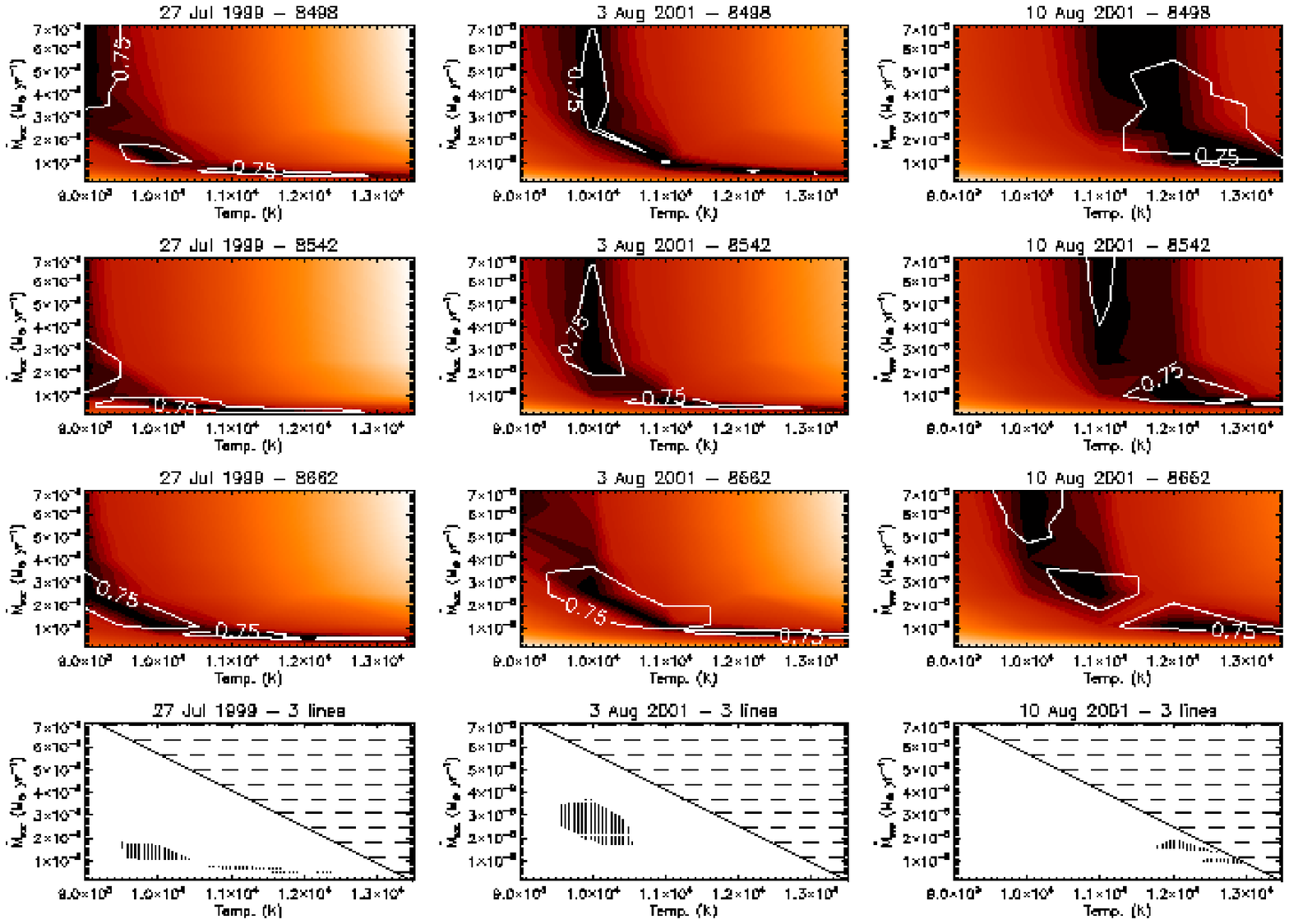}
\caption[Results from $\chi^2$ fits of the model to observations of DI Cep, and the confidence level contour, for varying mass accretion rate and maximum temperature of the gas.]{\label{fig:contours} Contours of $\chi^2$ values obtained from model fits to the observations of DI Cep, for varying mass accretion rate and maximum temperature of the gas. Darker areas represent better fits. The color/grayscale gradient is not linear, i.e., it its steeper near the minimum $\chi^2$. The white line indicates a confidence level that we set to the upper quartile of the $\chi^2$ probability distribution. Each column corresponds to one date of observation. In the last row we plot the intersection of the regions which lie above the confidence level on the three previous rows (dots). Extended features in the temperature axis direction are just a consequence of model lack of resolution. In the dashed region the continuum importance starts to be non-negligible. }
\end{figure*}
We compare our models to multi epoch observations of DI Cep carried out by GFP05 from 1998 to 2001. Further details on the observational procedures and instrumentation used are given in GFP05. The CaII IR triplet lines are blended with the Hydrogen Paschen series lines Pa 16-3 (8502.483\AA), Pa 15-3 (8545.382\AA) and Pa 13-3 (8665.018\AA), which are located at 157.1 km s$^{-1}$, 115.5 km s$^{-1}$ and 99.6 km s$^{-1}$, respectively, from the center of the CaII lines. There is no indication of their presence so we neglect them in our analysis. The presence of narrow components in the CaII IR triplet line profiles is also neglected because the narrow components are not seen. There were no veiling corrections made on the observed lines. GFP05 discussed the difficulty of estimating veiling in this wavelength region because of the nonexistence or weakness of photospheric absorption lines. However, some extrapolation can be done from shorter wavelengths leading us to conclude that the veiling is low, of the order of $\sim0.05$ to $\sim0.2$ and increasing slowly from 8498\AA\ to 8662\AA. In our normalized line profiles this implies that the lines can be stronger by 5 to 20 percent.

In general, the comparison between our model line profiles and observations should be carried out after removing from the latter the underlying photospheric component, i.e, after subtracting an appropriately veiled template spectrum. For the current DI~Cep observations this is not necessary since the emission lines are much stronger than the observed absorption strength on the template star.

DI Cep is a classical T Tauri star with spectral classification ranging from F4-K5. The most probable spectral type is between G5-G9 for a luminosity class V \citep{Ismailov2003}; \citet{Herbig1988} indicate a G8V spectral type. Using the tables of \citet{Hartmann1995} and the interval G5-G9V, the effective temperature is $5590\pm180$K. \citet{Kholopov1958} determines a distance to DI Cep of 300pc whereas \citet{Grinin1980} obtain a 200pc distance. We shall consider the range $250\pm50$pc in our analysis. Using photometry from \citet{Cohen1976} the average $<V-I>=1.41$ magnitudes. Using \citet{Hartmann1995} to deredden $V-I$ and \citet{Cardelli1989} extinction law ($R_{V}=3.1$) we derive $A_{V}=0.88\pm0.42$ magnitudes. Using 2MASS J band photometry and equation A1 from appendix A on \citet{Hartmann1995} to derive the bolometric magnitude, yields the luminosity $L_\star=3.8\pm1.8L_\odot$ and $R_\star=2.0\pm0.4R_\odot$. One should take these values as a mere estimate. Therefore we adopt the value $R_{*}=1.8R_\odot$ and $T_{eff}=5400K$ which results in a stellar mass of 1.42$M_\odot$, using the pre-main sequence evolutionary tracks of \citet{Siess2000}.

To determine the inclination $i$ we consider the measurement of the period and $V\sin i$ found in the literature. \citet{Bouvier1986} report $V\sin i=28$ km s$^{-1}$; \citet{Gameiro1993} found 23 Km s$^{-1}$, and in GFP05 $V\sin i$ is between 19 and 23 Km s$^{-1}$. We adopt $V\sin i=23.5\pm4.5$Km s$^{-1}$. \citet{Ismailov2003} found a DI Cep periodicity from the emission lines of 9.24 days and \citet{Kolotilov2004} 9.2 days from the U-V light curve. These spectroscopic measurements result in \hbox{$R_\star\sin i=4.3\pm0.8R_{\odot}$}, which largely exceeds $R_\star$ derived from photometry, and for the adopted stellar luminosity and using the same evolutionary tracks, $R_\star\sin i$ values imply a mass greater than $2M_\odot$. Given these problems, we adopt the value $i=60^o$, as the models match better the observations.

\begin{figure*}
\centering
\includegraphics[width=12cm]{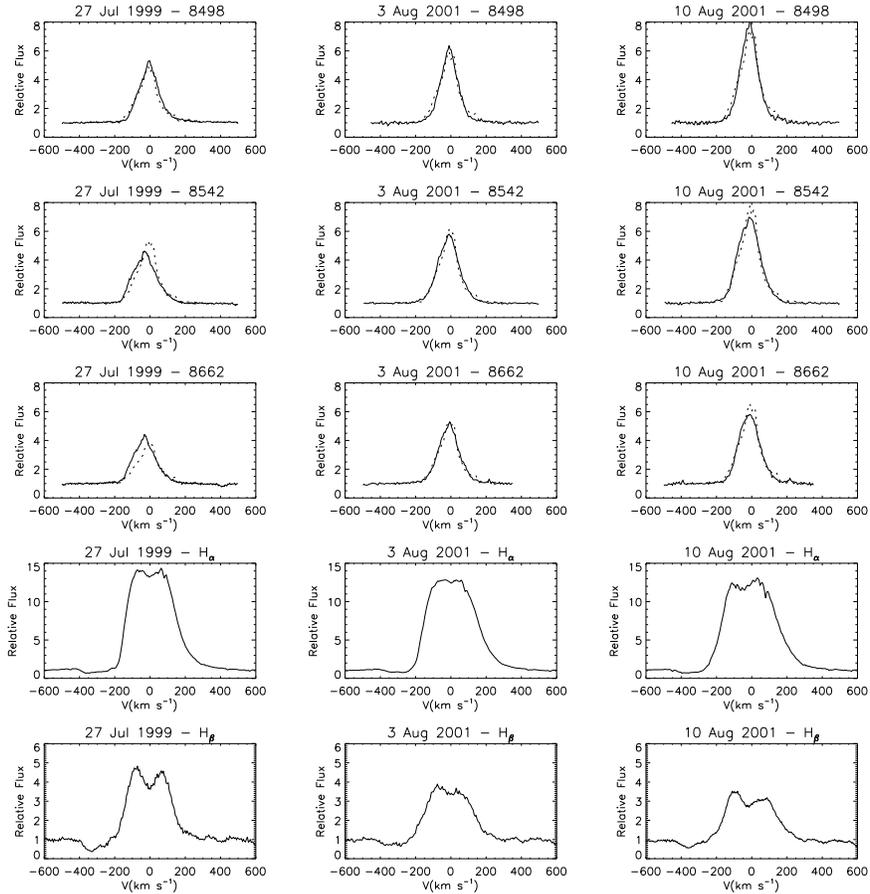}
\caption[Best fits of the model to DI Cep observations, in three different epochs for the IR triplet lines, and the $H_\alpha$ and $H_\beta$ observed profiles.]{\label{fig:dicepplots} Best fits of the model to DI Cep observations, in three different epochs for the IR triplet lines. The last two rows in the plot show the observed $H_\alpha$ and $H_\beta$ line profiles, for which no models were produced. We have used the gas temperature and mass accretion rate corresponding to the regions of Fig. \ref{fig:contours} bottom row. 27 July 1999: $\dot{M}_{acc}=8.0\times10^{-9}$M$_{\odot}$ yr$^{-1}$ and $T_{gas}=11000$K; 3 August 2001: $\dot{M}_{acc}=2.5\times10^{-8}$M$_{\odot}$ yr$^{-1}$  and $T_{gas}=10000$ K; 10 August 2001: $\dot{M}_{acc}=1.6\times10^{-8}$M$_{\odot}$ yr$^{-1}$ and  $T_{gas}=12100$K. Fixed parameters: $R_\star=1.8R_{\odot},\; M_\star=1.42M_{\odot},\; T_\star=5400K,\; T_r=8000K,\; Geometry=5.8-6R_\star,\; i=60^o$}
\end{figure*}

We are left with four parameters to set the runs, $T_{shock}$,$T_{gas}$,$\dot{M}_{acc}$ and the geometry. We have to make as much as plausible assumptions as we can to restrict the study to few parameters. Therefore, the temperature of the heated photosphere in the shock is assumed to be approximately 8000K which corresponds to the temperature derived from optical variability observations by \citet{Fernandez1996} and \citet{FernandezV1996}. After an initial exploration of the geometry effects on the line profiles,  we favoured a large and narrow magnetosphere since overall it produces  better fits to the DI Cep data. The profiles thus produced are narrower, more symmetric and exhibit more emission for lower mass accretion rates, in better agreement with the data. Independently of the values of $T_{shock}$, $T_{gas}$ and $\dot{M}_{acc}$, other geometries are never capable of fits as good as those obtained with a large and narrow geometry. In particular, the observed variability on DI Cep can never be explained in the context of our model by geometry changes even when accompanied by changing the remaining parameters. Afterwards, we tried to fine tune the mass accretion rate and the temperature of the gas such that the three IR triplet lines could be modeled simultaneously. To accomplish that we searched our grid of models for the minimum values of $\chi^2$, calculated from $F_{obs}$ (observed continuum normalized flux) and $F_{mod}$ (model continuum normalized flux). This analysis is shown in Fig. \ref{fig:contours}. We plot in white line the confidence level set to the upper quartile of the $\chi^2$ probability distribution, on three different dates of observation, with each row corresponding to fits of the 8498\AA, 8542 \AA\ and 8662 \AA\ lines. In the bottom, for each line, we plot the intersection of the regions bounded by the upper quartile threshold (dots). We also plot the $\chi^2$ contours in the color/grayscale gradient to illustrate the interplay between mass accretion rate and temperature. The contours for a given line show that the best fits can be produced with $\dot{M}_{acc}$ and $T_{gas}$ spanning a large range of values. (For example, $T_{gas}$ may decrease as long as $\dot{M}_{acc}$ increases accordingly, and the $\chi^2$ will remain constant). The darker vertical bands in the $\chi^2$ contours represent areas where an increase on $\dot{M}_{acc}$ does not produce an increase in line emission, keeping $\chi^2$ constant for the same $T_{gas}$. The model profiles are saturated which means that the lines have completely thermalized. However, as the mass accretion rate increases, the assumption of a continuum optically thin magnetosphere will eventually become invalid, and the optical photospheric spectrum would fade away being swamped by an ever increasing continuum emission, putting an upper boundary in the $\chi^2$ vertical bands. For mass accretion rates as low as $3\times10^{-8}$M$_{\odot}$ yr$^{-1}$ in Fig. \ref{fig:contours}, at a gas temperature of 12000K, we have to start considering the effects from the Paschen continuum emission. For a gas temperature of 9000K the magnetosphere starts to have non-negligible contributions towards the continuum emission for $\dot{M}_{acc}$ starting at about $7.5\times10^{-8}$M$_{\odot}$ yr$^{-1}$. The dashed lines on Fig. \ref{fig:contours} mark the regions where this magnetospheric Paschen continuum becomes important. We remark that these regions are valid if the accretion geometry was approximately axisymmetric as in our model. For a non-axisymmetric geometry, with the same $T_{gas}$ and same column density, higher mass accretion rates are possible since there are fewer accretion columns emitting, and the photosphere is also more exposed.
 
 Using the region of the bottom row in Fig. \ref{fig:contours} we made fits to the lines on three different observational dates (see Fig. \ref{fig:dicepplots}), illustrating the generally good agreement between model and observed line properties.
 
\begin{figure}
	\centering
	\resizebox{\hsize}{!}{\includegraphics{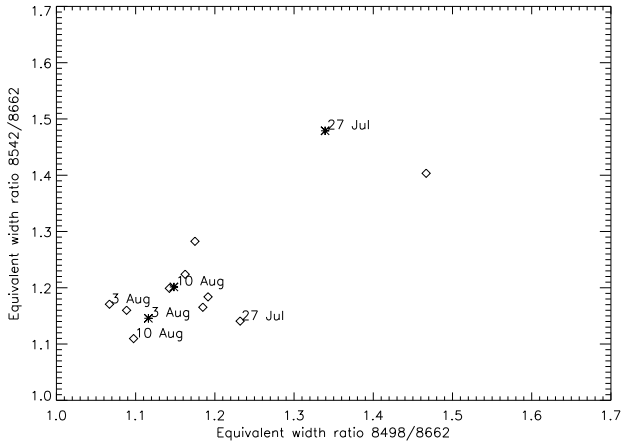}}
	\caption{\label{fig:lineratios} Equivalent width ratios of the CaII IR triplet lines (diamonds) from \citet{Gameiro2004} observational data sample of DI Cep. Equivalent width line ratios for models that represent the 27 July 1999, 3 August 2001 and 10 Aug 2001 data (Fig. \ref{fig:dicepplots}) are shown as asterisks.}
\end{figure}

Figure \ref{fig:lineratios} shows the equivalent width ratios for all the observations of GFP05 (diamonds). Model equivalent width ratios obtained from the fits of Fig. \ref{fig:dicepplots}, are plotted on Fig. \ref{fig:lineratios} as asterisks.

\section{Discussion}

 The fits to the observations in Fig. \ref{fig:dicepplots} are very acceptable, in particular given the approximations we are using. Magnetospheric infall can therefore be the most important process for the CaII IR triplet broad component line formation. The Hydrogen profiles show P~Cygni character, which arises in a wind. In addition, the hydrogen lines are much broader than the CaII lines, which may also be accounted for by wind emission. Our magnetospheric accretion model is not able to tackle this wind emission, and hence the hydrogen line profiles would never be reproduced in this context.
 
It is possible to model the observed variability by varying $\dot{M}_{acc}$ and temperature of the gas (Fig. \ref{fig:dicepplots}) simultaneously. To further test this conclusion one could do observations/modelling of other elements. One could also compare observed absolute line fluxes (not available with the dataset discussed here) with absolute line fluxes resulting from the models.
 
From the results of Fig. \ref{fig:contours}, and looking at an individual line, we can accommodate mass accretion rates ranging from $5\times10^{-9}$M$_{\odot}$ yr$^{-1}$ to $7.5\times10^{-8}$M$_{\odot}$ yr$^{-1}$ for a gas temperature of 9000K, and a maximum $\dot{M}_{acc}$ of the order of $3\times10^{-8}$M$_{\odot}$ yr$^{-1}$ at 12000K, for an individual line. Simultaenous good fits of the three CaII lines for each date of observation enable us to restrict the gas temperature and the mass accretion rate to narrow intervals. The variations we find in the gas temperatures and mass accretion rates from one date to another lie within $10000-12000$K and $7\times10^{-9}-4\times10^{-8}$M$_{\odot}$ yr$^{-1}$, respectively. Taking an extreme upper limit on the accretion luminosity as $L_{acc}\approx GM_{*}\dot{M}_{acc}/R_{*}\approx L_{*}$ then for the stellar parameters that we are assuming, the mass accretion rate must be less than $\dot{M}_{acc}\lesssim1\times10^{-7}M_{\odot}/yr$, otherwise $L_{acc}$ would be much higher than $L_{*}$. Our values lie comfortably below this limit.
 
 \citet{Fernandez1996}, find that $\dot{M}_{acc}\geq6\times10^{-9}$M$_{\odot}$ yr$^{-1}$ assuming that all the energy dumped in the UV continuum and lines is due to kinetic energy released from infall from a corotation radius of 8.3 stellar radii. \citet{Castro1999} used the predictions of the semiforbidden line ratios SiIII/OIII and SiIII/CIII for the accretion shock (from models by \citet{Lamzin1998}), the observed semiforbidden line ratios, and the more uncertain values of extinction and distance values to DI Cep, to obtain a value of mass accretion rate $\dot{M}_{acc}=1.9\times10^{-7}$M$_{\odot}$ yr$^{-1}$. They assumed the distance of \citet{Kholopov1958}, i.e., 300pc and $A_V=0.24$. Using equation 4 of their work and a distance of 200pc from \citet{Grinin1980} would bring $\dot{M}_{acc}$ to about $8.4\times10^{-8}$M$_{\odot}$ yr$^{-1}$; however, using our estimate of $A_{V}$ one would get $1.3\times10^{-7}\lesssim\dot{M}_{acc}\lesssim2.9\times10^{-6}$M$_{\odot}$ yr$^{-1}$, a value which results in an accretion luminosity far too high.
 
\citet{Krull2000a} find $\dot{M}_{acc\,1}=4.0\times10^{-7}M_{\odot}yr^{-1}$ and $\dot{M}_{acc\,2}=7.8\times10^{-6}M_{\odot}yr^{-1}$, using two empirical relations between the $CIV 1549$\AA\ line excess luminosity with $\dot{M}_{acc}$, values that would imply, again, an optical continuum veiling much higher than observed. We could not determine the influence of using our stellar parameters in their work. However, there are important sources of uncertainty in their determinations of mass accretion rates. In particular, the empirical relations used in these determinations are strongly dependent on the extinction corrections assumed for the stars upon which the relations are derived. In addition, the empirical relations rely on \citet{Hartigan1995} determinations of mass accretion rates. These are based on the boundary layer hypothesis which overestimate the mass accretion rate compared to a magnetospheric accretion model.
 
 Finally, GFP05 find a projected mass accretion rate $\dot{M}_{acc}\cos(i_{to})=2.5\times10^{-7}$M$_{\odot}$ yr$^{-1}$ using fits of the models of \citet{Calvet1998} to their observed wavelength veiling dependence in the spectral range $4000\AA-8000\AA$. Here, $i_{to}$ is the angle between the accretion tube and the observer direction. If we use the stellar mass and radius applied in our work the projected mass accretion rate would be $\dot{M}_{acc}\cos(i_{to})=6.3\times10^{-8}$M$_{\odot}$ yr$^{-1}$. As the inclination $i_{to}$ is not known, the mass accretion rate could increase by the factor $1/\cos(i_{to})$.
  
  Given this discussion, it is evident that there is a large uncertainty in the mass accretion rate. The values that we find are consistent with the lower limit derived by \citet{Fernandez1996} and lower than the value found in GFP05. The other determinations mentioned are likely to be overestimating DI Cep's mass accretion rate.

The observed CaII line peaks in DI Cep follow the relation I$^{p}_{8498 \AA} >$ I$^{p}_{8542\AA} >$ I$^{p}_{8662\AA}$ (Fig. \ref{fig:dicepplots}). However, because of the narrowness of the 8498\AA\ line, the equivalent width ratios follow the tendency F$_{8498\AA} \approx$ F$_{8542\AA} >$ F$_{8662\AA}$ (Fig. \ref{fig:lineratios}). Figure \ref{fig:lineratios} also puts in evidence that when the 8498\AA\ line becomes stronger the others do not have a proportional increase. 

The models reproduce well the line peaks but predict a 8542\AA\ line peak stronger than the other two IR triplet lines (Fig. \ref{fig:dicepplots}). Nevertheless, the more common observed peak pattern I$^{p}_{8498\AA} >$ I$^{p}_{8662\AA}$ is reproduced. The lines have different intensities mainly due to the different source functions (see figure \ref{fig:threesources}), which follow the relation of the collisional cross sections ($\sigma^{c}$) used, i.e., $\sigma^{c}_{8542\AA} > \sigma^{c}_{8498 \AA} > \sigma^{c}_{8662\AA}$. Therefore, the peak ratios are sensitive to the accuracy of the latter, which for example in \citet{Shine1974} are different from those we use. As we are exclusively considering the extended Sobolev approximation, and since the line peaks are formed where the approximation breaks down, i.e., where the material lifts off from the disk, our model results around the line center should be taken cautiously.

Figure \ref{fig:lineratios} shows that the model equivalent width ratios are of the same order as the observed ones with the exception of the more problematic night of 27th of July, 1999. However, the variability of the equivalent width ratios does not have a complete correspondence in the model. Our variations on $\dot{M}_{acc}$ and $T_{gas}$ alone, do not totally produce the necessary differential changes in the profiles to explain their variability completely.

 In some CTTS, specially those with low mass accretion rates, the broad component starts to fade away and a narrow component becomes prominent. From the \citet{Muzerolle1998} sample we see that the three stars with NC type CaII profiles have mass accretion rates of about $3.4\times10^{-9}$M$_{\odot}$ yr$^{-1}$, which might be interpreted as a lower limit on the mass accretion for the disappearance of the BC ($\dot{M}_{min}$). In our models, depending on the used geometry and temperature of the gas, we find in general that $\dot{M}_{min}$ should be between $1\times10^{-9}$ to $5\times10^{-9}$ M$_{\odot}$ yr$^{-1}$, in agreement with the above lower limit.
 
 An important remark is that \citet{Muzerolle1998} found a correlation between the IR triplet line total flux 8542\AA\ and the mass accretion rate, for mass accretion rates between $10^{-9}$M$_{\odot}$ yr$^{-1}$ and $10^{-6}$M$_{\odot}$ yr$^{-1}$. Note that this relation also includes stars with NC profiles. Therefore, not only the relation should be explained in the broad component regime by our models but also the smooth transition between the broad component and the narrow component regime should be accounted for. This seems to indicate that a multi component model is necessary to predict narrow component fluxes coming either from a hot chromosphere, accretion shock or wind, and the fluxes from the BC formed in the magnetospheric accretion flow. \citet{Subu2005} have extended the relation of the CaII lines fluxes with mass accretion rates to the brown dwarfs. Here $\dot{M}_{acc}$ can be as low as $\dot{M}_{acc}=1\times10^{-11}$ M$_{\odot}$ yr$^{-1}$. Such a multi origin component model for CTTS should also be able to reproduce the correlation in this low mass regime.

\section{Conclusions}

The magnetospheric accretion model can explain well the CaII IR line profiles of DI Cep, as well as the observed variability, using slightly different mass accretion rates and different gas temperatures for different epochs. The mass accretion rates found for DI Cep that fit the observations are between $5\times10^{-9}$ M$_{\odot}$ yr$^{-1}$ and $2\times10^{-8}$ M$_{\odot}$ yr$^{-1}$. The peak ratios are not perfectly reproduced, although the models can predict the observed ratio I$^{p}_{8498\AA} >$ I$^{p}_{8662\AA}$. It is difficult to fully interpret the observed variability using variations of $\dot{M}_{acc}$ and $T_{gas}$. Further tests of the model and restrictions on the parameter space can be done with simultaneous observations/modeling of other elements and by trying to reproduce the relation between the mass accretion rates and the CaII infrared line fluxes.

\begin{acknowledgements}
We would like to thank E. Avrett and J.Hernandez for their useful help. R. Azevedo acknowledges support by Funda\c{c}\~{a}o para a Ci\^{e}ncia e Tecnologia, through grant BD/14164/2003. F. Gameiro, D. Folha and R. Azevedo also acknowledge support by grants POCTI/1999/FIS/34549 and POCI/CTE-AST/55691/2004 approved by FCT and POCTI, with funds from the European Community programme FEDER.

\end{acknowledgements}

\bibliography{bibfile}

\end{document}